\documentclass[12pt]{article}
\usepackage{a41}

\newcommand\GeV{\,\mbox{GeV}}

\begin{document}
\thispagestyle{empty}

\mbox{}
\begin{flushleft}
DESY  02--193 \hfill {\tt hep-ph/0211191}\\
November 2002\\
\end{flushleft}
\vspace*{\fill}
\begin{center}
{\LARGE\bf 
\boldmath{${O}(\alpha^2 L)$} Radiative Corrections to}

\vspace{2mm}
{\LARGE\bf Deep Inelastic \boldmath{$ep$} Scattering}

\vspace*{20mm}
\large
\begin{tabular}[t]{c}
Johannes Bl\"umlein and Hiroyuki Kawamura
\\

\vspace{2em}
\\

{\it Deutsches Elektronen--Synchrotron, DESY, }
\\
{\it Platanenallee 6,
D--15735 Zeuthen,  Germany}
\\
\end{tabular}
\end{center}
\vspace*{\fill}
\begin{abstract}
\noindent
The leptonic QED radiative corrections are calculated in the next-to-leading 
log approximation  ${\cal O}\left[\alpha^2 \ln(Q^2/m_e^2)\right]$ for 
unpolarized deeply inelastic $ep$--scattering in the case of mixed variables. 
The corrections are determined using mass factorization in the OMS--scheme
for the double--differential scattering cross sections.
\end{abstract}
\vspace*{\fill}
\newpage
\noindent
\section{Introduction}

\vspace{1mm}
\noindent
Deeply inelastic electron--nucleon scattering provides one of the 
cleanest methods to test QCD by measuring the scaling violations of
structure functions from which the QCD--scale $\Lambda_{\rm QCD}$, one 
of the fundamental parameters of the Standard Model, is derived with high
precision. The QCD--analysis also delivers the various twist--2 parton 
distributions as functions of $x$ and $Q^2$, the detailed knowledge of 
which is necessary for the forthcoming experimental search for the 
Higgs--boson and new, yet undiscovered states beyond the Standard Model, 
at TEVATRON and LHC. One of the major goals of the experiments H1 and ZEUS 
at the $ep$--collider HERA at DESY is to perform a QCD test at large 
space--like virtualities $Q^2$ at high precision. This presumes to know 
the QED radiative corrections to the double--differential scattering cross 
sections of deeply inelastic $ep$ corrections as precisely as possible.
Previous calculations of the radiative corrections for the unpolarized 
cross sections at leading order [1--5]\footnote{For QED corrections
to polarized lepton scattering off polarized nucleons see~\cite{POLA}.},
the leading--log level [7--12] to leading and higher orders, and
QED--resummations of small--$x$ terms~[11,13]\footnote{For related
numerical results for QCD singlet evolution see \cite{SING}.} orders
revealed that these corrections are very large, particularly in the 
interesting regions of small $x$ and for almost all $x$ at large 
inelasticities $y=Q^2/(Sx)$. The corrections are, moreover, complicated
by a new type of sizeable contributions as the Compton--peak~\cite{COMPT}.
This makes it necessary to extend the calculations to higher orders.

The higher order leading--logarithmic contributions $O\left[(\alpha L)^k
\right]$ to QED corrections are obtained as the leading order solution
of the associated renormalization group equations~\cite{CS} for mass
factorization. These corrections are universal, process--independent
w.r.t. their structure, and are given in terms of Mellin--convolutions
of leading order QED splitting functions. The next--to--leading order 
(NLO) corrections can be obtained along the same line. However, besides 
the splitting functions to NLO also the respective process--dependent
Wilson coefficients and operator matrix elements in the on-mass-shell
(OMS) scheme contribute. In the past this method was applied to a single
differential distribution for the $O(\alpha^2)$ initial--state QED
radiative corrections to $e^+e^- \rightarrow \mu^+\mu^-$ in 
Ref.~\cite{BBN}.

In this letter we calculate the leptonic QED corrections to deeply
inelastic $ep$ scattering, where we define the double--differential
scattering cross section choosing mixed variables~[8,2,5], in
$O(\alpha^2 L)$. If compared to the case dealt with in \cite{BBN} to
$O(\alpha^2 L)$ the present calculation is more complicated due to the
emergence of final state radiation, the double--differential cross section
and the relevant rescaling, which implies non--Mellin type convolutions 
in general. After summarizing main kinematic aspects we present the 
different contributions to the leptonic NLO QED corrections. Further 
details of the calculation and numerical results will be given
elsewhere~\cite{JBHK3}.
\section{Mixed variables}

\vspace{1mm}
\noindent
$ep$ collider experiments allow to measure the kinematic variables
defining the inclusive deep--inelastic scattering cross sections in 
various ways since four kinematic variables are available in principle
with the energies and angles of both the outgoing lepton and the 
struck--quark, see e.g.~\cite{KIN}. At the Born level all methods are
equivalent, however, resolution effects as a consequence of the
detector's structure, differ in certain kinematic regions. The 
Bremsstrahlung--effects of QED radiative corrections change this picture 
drastically and the QED correction factors depend on the way the kinematic
variables, as e.g. Bjorken--$y$ and the virtuality $Q^2$ are 
measured.\footnote{See e.g. Refs.~[8,2,5] for a comparison of a wide
range of different choices of measurement.} In the present paper
we calculate the NLO radiative corrections in the case of neutral
current deep--inelastic scattering for mixed variables, i.e. that 
$Q^2 = Q^2_l$ is measured at the leptonic and $y=y_h$ is measured at the 
hadronic vertex, and $x_m =Q^2_l/(S y_h)$.  The Born cross
section for $\gamma$--exchange is given by\footnote{As we discuss 
radiative corrections we will speak about an $O(\alpha^k)$ contribution 
for a cross section being proportional to $\alpha^{2+k}$.}~:
\begin{equation}
\frac{d^2 \sigma^{(0)}}{dy dQ^2} \equiv C^{(0,0)}(y,Q^2)
= \frac{2\pi \alpha^2}{y Q^4} \left[
y^2~2xF_1(x,Q^2) + 2(1-y)~F_2(x,Q^2)\right],
\end{equation}
with
\begin{eqnarray}
F_1(x,Q^2) &=& \frac{1}{2} 
\sum_{k=1}^{N_f} \left[q_k(x,Q^2) + \overline{q}_k(x,Q^2)
\right],\\
F_2(x,Q^2) &=& 2x F_1(x,Q^2) + F_L(x,Q^2)~.
\end{eqnarray}
Here, $F_{1,2,L}(x,Q^2)$ denote the nucleon structure functions for
photon exchange, and $q(x,Q^2)$ and $\overline{q}(x,Q^2)$ are
the quark-- and antiquark distribution functions. The sub--system 
variables obey the following rescaling relations for initial-- and final--state 
radiation~:
\begin{eqnarray}
\label{eqRESC1}
{\sf ISR~:}& &~~~\widehat{y} = \frac{y_h}{z},~~~\widehat{Q}^2 
           = z Q^2_l = Q_h^2,~~~\widehat{S} 
           = zS,~~~\widehat{x} = zx_m = x_h,
           \nonumber \\
& &~~~J^I(z) = 1,~~~~~z_0^I = {\rm min}\left\{y_h,\frac{Q_0^2}{Q_l^2}
\right\}~,  \\
\label{eqRESC2}
{\sf FSR~:}& &~~~\widehat{y} = y_h,~~~\widehat{Q}^2
           = \frac{ Q^2_l}{z} = Q^2_h,~~~\widehat{S} 
           = S,~~~\widehat{x} = \frac{x_m}{z} = x_h,
           \nonumber \\
& &~~~J^F(z) = \frac{1}{z},~~~~~z_0^I = x_m~.
\end{eqnarray}
Here, $J^{I,F}(z)$ are the initial-- and final--state Jacobians
$d^2(\widehat{y},\widehat{Q}^2)/d^2(y_h,Q^2_l)$, and $z_0$ marks the
lower bound of the sub--system rescaling variable $z~\epsilon~[z_0,1]$.
The rescaling in Eqs.~(\ref{eqRESC1},\ref{eqRESC2}) was chosen such that
both the initial-- and final--state operator matrix elements can be
expressed by a variable $z~\epsilon~[0,1]$. $Q_0^2$ is introduced as
a scale to cut away contributions of the Compton peak. Although these
terms do formally belong to the QED radiative corrections, they stem
from a kinematic domain of low virtualities and are therefore not being
associated to deep inelastic scattering. The scale $Q_0^2 \simeq 1~\GeV^2$ 
or larger can be chosen
by experiment accepting only those events in the sample to be analyzed
for which the {\it hadronic} $Q^2$ is larger than $Q_0^2$. By this measure
the Compton peak is cut away widely and the QED--correction factor is
dominated by a deep--inelastic sub--process by far. Finally we note that 
in the case of mixed variables the leptonic QED radiative corrections can
be easily grouped into those for the initial and final state. 
The separation scale between the two kinematic regions is $Q_l^2$, i.e.
we speak of initial state radiation for $Q_h^2 < Q_l^2$ and of final
state radiation for $Q_h^2 > Q_l^2$.

\section{NLO corrections}

\vspace{1mm}
\noindent
In this paper we limit the consideration to the calculation of the
NLO corrections to leptonic variables for one--photon exchange in
electron--nucleon scattering. This approach is widely model independent
and allows to refer to general non--perturbative parameterizations of
the structure functions which describe the hadronic tensor. In this way
a direct unfolding of the experimentally measured structure functions is
possible down to the range in $Q^2$ and $x$ in
which partonic approaches fail 
to provide a description of structure functions.\footnote{Quarkonic and
lepton-hadron interference contributions are not included in this 
approach. They are of smaller size in general~\cite{QEDQ1,LLA1,QEDQ2}.}
The radiative corrections are thus valid as well for inclusive
{\it diffractive} $ep$--scattering, see e.g.~\cite{DIFFR}.
The radiative corrections to be calculated concern a neutral--current 
process in which the lepton stays the same~: both an incoming electron
(positron) results into an outgoing electron (positron). As was already
pointed out in Ref.~\cite{LLA1} changes of the fermion species are
possible starting with 2--loop order. To keep the QED $K$--factor close
to the basic process we limit the consideration of the present paper
to those contributions in which the electron--type is preserved. The 
scattering cross section to $O(\alpha^2)$ reads~:
\begin{eqnarray}
\label{eq2L}
\frac{d^2 \sigma}{dy_h dQ_l^2} &=&
C^{(0,0)}(y_h,Q^2_l) + \frac{\alpha}{2\pi} \left\{
\ln\left(\frac{Q_l^2}{m_e^2}\right) C^{(1,0)}(y_h,Q_l^2)+
C^{(1,1)}(y_h,Q_l^2) \right\}\\ & &
+ \left(\frac{\alpha}{2\pi}\right)^2
\left\{
\ln^2\left(\frac{Q_l^2}{m_e^2}\right) C^{(2,0)}(y_h,Q_l^2)+
\ln\left(\frac{Q_l^2}{m_e^2}\right) C^{(2,1)}(y_h,Q_l^2)+
C^{(2,2)}(y_h,Q_l^2) \right\} \nonumber
\end{eqnarray}
with $C^{(0,0)}(y_h,Q_l^2) ={d^2 \sigma^0}/{dy_h dQ_l^2}$ and
$\alpha = e^2/(4\pi)$.
The  $O\left[(\alpha L)\right]$ and $O\left[(\alpha L)^2\right]$ 
corrections were calculated in Ref.~\cite{JB94}. The term 
$C^{(1,1)}(y_h,Q_l^2)$ was derived in Ref.~\cite{ABKR} 
completing the $O(\alpha)$ corrections. We re-calculated 
these corrections and agree with the previous results.

The NLO--correction $C^{(2,1)}(y_h,Q_l^2)$ can be obtained representing 
the scattering cross section using mass--factorization, see 
\cite{BBN}\footnote{Very recently also the electron energy spectrum
in muon decay has been calculated using this method~\cite{AM}.}.
Although the differential 
scattering cross section does not contain any mass singularity, one may 
decompose it in terms of Wilson coefficients and operator-matrix elements
being convoluted with the Born cross section.
In this decomposition both the operator matrix elements and the Wilson
coefficients depend on the factorization scale $\mu^2$. 
One writes the scattering
cross section as
\begin{eqnarray}
\label{eqCONV}
\frac{d^2 \sigma}{dy_h dQ_l^2} &=&
\frac{d^2 \sigma^0}{dy_h dQ_l^2} \otimes \Biggl\{ 
  \Gamma^I_{ee} \otimes \hat{\sigma}_{ee} \otimes \Gamma^F_{ee}
+ \Gamma^I_{\gamma e} \otimes \hat{\sigma}_{e\gamma} \otimes \Gamma^F_{ee}
+ \Gamma^I_{ee} \otimes \hat{\sigma}_{\gamma e} \otimes 
\Gamma^F_{e \gamma}
\Biggr\}~,
\end{eqnarray}
with $\Gamma^{I,F}_{ij}(z,\mu^2/m^2_e)$ the initial and final state
operator matrix elements and $\hat{\sigma}_{kl}(z,Q^2/\mu^2)$ the 
respective
Wilson coefficients. $\otimes$ denotes a convolution, which depends
on specific rescalings of the chosen kinematic variables for the
differential cross sections and is specified below. Both the
operator matrix elements and the Wilson coefficients obey the
representations
\begin{eqnarray}
\Gamma^{I,F}_{ij}\left(z,\frac{\mu^2}{m_e^2}\right) &=&
\delta(1-z) + \sum_{m=1}^{\infty} \left (\frac{\alpha}{2\pi} \right)^m
\sum_{n=0}^m
\Gamma_{ij}^{I,F(m,n)}(z)
\ln^{m-n}\left(\frac{\mu^2}{m_e^2}\right)~, \\
\hat{\sigma}_{kl}\left(z,\frac{Q^2}{\mu^2}\right) &=&
\delta(1-z) + \sum_{m=1}^{\infty} \left(\frac{\alpha}{2\pi} \right)^m
\sum_{n=0}^m 
\widehat{\sigma}_{kl}^{(m,n)}(z)
\ln^{m-n}\left(\frac{Q^2}{\mu^2}\right)~.
\end{eqnarray}
The sequences $\{ij\}$ and $\{kl\}$ in the above do always denote 
$j(l)$ for the incoming and $i(k)$ the outgoing particle. As the 
differential cross section is $\mu$--independent, the cross section is 
expressed by  convolutions
of the functions $\Gamma_{ij}^{I,F(m,n)}(z)$ and 
$\widehat{\sigma}_{kl}^{(m,n)}(z)$ such that the $\mu^2$--dependence 
cancels and a structure like  Eq.~(\ref{eq2L}) is obtained. The 
treatment in the OMS scheme assumes that the light fermion mass, $m_e$, 
is kept everywhere it is giving a finite contribution to the scattering
cross section if compared to the large scale $Q^2$, i.e. the only terms 
being neglected are power corrections which are of 
$O\left[(m_e^2/Q^2)^k\right],~k \geq 1$ and therefore small. 
The last step is necessary to maintain the anticipated convolution 
structure which, in parts, is of the Mellin--type, as also in
a massless approach.

In the subsequent relations we make frequent use of the rescaling 
(\ref{eqRESC1},\ref{eqRESC2}). For this purpose we introduce the
following short--hand notation for the rescaling of a function $F(y,Q^2)$
\begin{eqnarray}
\label{RESCA}
\widetilde{F}_{I,F}(y,Q^2) 
= F\left(y=\widehat{y}_{I,F},Q^2=\widehat{Q}^2_{I,F}\right)~,
\end{eqnarray}
where $I,F$ label the respective type of rescaling.
The NLO--corrections may be grouped into the following contributions~:
 
\vspace{2mm}\noindent
{\sf
\renewcommand{\labelenumi}{\theenumi}
\begin{enumerate}
\item[i]~~LO initial and  final state radiation off 
$C^{(1,1)}_{ee}(y,Q^2)$
\item[ii]~~coupling constant renormalization of 
$C^{(1,1)}_{ee}(y,Q^2)$
\item[iii]~~LO initial state splitting of $P_{\gamma e}$ 
at $C^{(1,1)}_{e \gamma}(y,Q^2)$
\item[iv]~~LO final   state splitting of $P_{e \gamma}$ at 
$C^{(1,1)}_{\gamma e}(y,Q^2)$
\item[v]~~NLO initial and  final state radiation off 
$C^{(0,0)}_{ee}(y,Q^2)$
\end{enumerate}
}

\vspace{1mm}
\noindent
The function $C^{(2,1)}(y,Q^2)$ then is given by
\begin{eqnarray}
\label{C2SUM}
C^{(2,1)}(y,Q^2) = \sum_{i = {\sf i}}^{\sf v} C_i^{(2,1)}(y,Q^2),
\end{eqnarray}
where $y=y_h$, $Q^2=Q^2_l$.
The function $C^{(2,1)}_{\sf i}(y,Q^2)$ is obtained by
\begin{eqnarray}
\label{C21i}
C^{(2,1)}_{\sf i}(y,Q^2) &=& \int_0^1 dz P_{ee}^0(z) \left[\theta\left
(z-z_0^I\right) J^I(z) \widetilde{C}^{(1,1)}_I(y,Q^2)
-{C}^{(1,1)}(y,Q^2) \right] \nonumber\\
&+&
\int_0^1 dz P_{ee}^0(z) \left[\theta\left(z-z_0^F\right) 
J^F(z) \widetilde{C}^{(1,1)}_F(y,Q^2)
-{C}^{(1,1)}(y,Q^2) \right],
\end{eqnarray}
where $C^{(1,1)}(y,Q^2)$ denotes the non--logarithmic part of the
$O(\alpha)$ correction~\cite{ABKR,JBHK3} and $P_{ee}^0(z)$ is the
LO fermion--fermion splitting function
\begin{eqnarray}
\label{P0ee}
P_{ee}^0(z)      = \frac{1+z^2}{1-z}~.
\end{eqnarray}
To illustrate the convolution structure in (\ref{eqCONV}) we note that
\begin{eqnarray}
C^{(1,1)}(y,Q^2) = \left[\widehat{\sigma}^{(1,1)}_{ee}
+ \Gamma_{ee}^{I,(1,1)} + \Gamma_{ee}^{F,(1,1)} \right] \otimes
C^{(0,0)}(y,Q^2)~.
\end{eqnarray}

Also the LO off-diagonal splitting functions
\begin{eqnarray}
\label{P0OD}
P_{e\gamma}^0(z) &=& z^2+(1-z)^2 \\
P_{\gamma e}^0(z)&=& \frac{1+(1-z)^2}{z}
\end{eqnarray}
occur in other contributions to $C^{(2,1)}$. Here both for LO and NLO
splitting functions for equal particle transitions we write the 
contributions for $z < 1$ and account for the +-functions in explicit 
form below.

We express the final result for $C^{(2,1)}(y,Q^2)$ in terms of the
QED coupling constant $a(m_e^2)$. Therefore we have to expand all
contributions using
\begin{eqnarray}
\label{RUN}
\alpha(\mu^2) = \alpha(m_e^2) 
\left[1 - \frac{\alpha}{4\pi} 
\beta_0 \ln\left(\frac{\mu^2}{m_e^2}\right) \right]~,
\end{eqnarray}
with $\beta_0 = -4/3$. In the present paper we only consider photonic
corrections and those due to a single light fermion. The generalization
to various fermions and non--perturbative contributions due to light 
quarks is, however, straightforward.
The running--coupling contribution to $C^{(2,1)}(y,Q^2)$ resulting from
$C^{(1,1)}(y,Q^2)$ is thus given by
\begin{eqnarray}
\label{C2ii}
C^{(2,1)}_{\sf ii}(y,Q^2) =  - \frac{\beta_0}{2} C^{(1,1)}(y,Q^2)~. 
\end{eqnarray}

The contributions $C^{(2,1)}_{\sf iii,iv}(y,Q^2)$ refer to two new
$O(\alpha)$ cross sections~: $d^2 \sigma^{\gamma e,(1)}/dydQ^2$
and $d^2 \sigma^{e \gamma,(1)}/dydQ^2$. The corrections are
\begin{eqnarray}
\label{C2iii}
C^{(2,1)}_{\sf iii}(y,Q^2) &=&  \int_{z_0^I}^1 dz P_{\gamma e}^0(z)
J^I(z) \widetilde{C}^{(1,1)}_{e\gamma,I}(y,Q^2)~, \\
\label{C2iv}
C^{(2,1)}_{\sf iv}(y,Q^2) &=&  \int_{z_0^F}^1 dz P_{e \gamma}^0(z)
J^F(z) \widetilde{C}^{(1,1)}_{\gamma e,F}(y,Q^2)~.
\end{eqnarray}
In these processes electrons and positrons are pair produced in the final
state. The $O(\alpha)$ sub--system cross sections read~: 
\begin{eqnarray}
\frac{d^2 \sigma^{\gamma e,(1)}}{dydQ^2} &=&
\frac{\alpha}{2\pi}\left[\ln\left(\frac{Q^2}{m_e^2}\right) 
C^{(1,0)}_{\gamma e}(y,Q^2)
+  C^{(1,1)}_{\gamma e}(y,Q^2)\right]~,\\
\frac{d^2 \sigma^{e \gamma,(1)}}{dydQ^2} &=&
\frac{\alpha}{2\pi}\left[\ln\left(\frac{Q^2}{m_e^2}\right) 
C^{(1,0)}_{e \gamma}(y,Q^2)
+  C^{(1,1)}_{e \gamma}(y,Q^2)\right]~.
\end{eqnarray}
Here, the functions $C^{(1,0)}_{ij}(y,Q^2)$ are given by
\begin{eqnarray}
C_{e\gamma}^{(1,0)}(y,Q^2) &=& \int_{z_0^I}^1 dz P_{e\gamma}^0(z) J_I(z)
\widetilde{C}^{(0,0)}_I(y,Q^2)~,\\
C_{\gamma e}^{(1,0)}(y,Q^2) &=& \int_{z_0^F}^1 dz P_{\gamma e}^0(z) J_F(z)
\widetilde{C}^{(0,0)}_F(y,Q^2)~,
\end{eqnarray}
and  the functions $C^{(1,1)}_{e\gamma}(y,Q^2)$ and 
$C^{(1,1)}_{\gamma e}(y,Q^2)$ are~:
\begin{eqnarray}
C_{e\gamma}^{(1,1)}(y,Q^2) &=&
\frac{2\pi \alpha^2}{y_h Q_l^4} \int_{z_0^I}^1 \frac{dz}{z^2} \Biggl\{
\Biggl[ P_{e\gamma}^0(z) \ln\left[\frac{(1-z)(z-y_h)}{z}\right]
-(1+2z) \ln\left(\frac{z}{y_h}\right) \nonumber\\ & &
~~~~~~~~~~~~~~~~-1  + \frac{z}{y_h} + 2z(1-z)
\Biggr]~y_h^2~2x_m F_1(x_h,Q^2_h) \nonumber\\
& &
~~~~~~~~~~~~~~~~+2\Biggl[(z-y_h) P_{e\gamma}^0(z) 
\ln\left[\frac{(1-z)(z-y_h)}{z}
\right] - y_h \ln\left(\frac{z}{y_h}\right) \nonumber\\
& &
~~~~~~~~~~~~~~~~+ (z-y_h) \left[
\frac{y_h}{z} - 1 -2y_h  +8z + 2y_hz - 8z^2\right]\Biggr]
F_2(x_h,Q^2_h)\Biggr\}  \nonumber
\\
& & +
\frac{2\pi \alpha^2}{y_h Q_l^4} \int_{z_0^F}^1 dz \Biggl\{\Biggl[
\frac{1}{z} P_{\gamma e}^0(z) \ln\left(\frac{1-y_h z}{1-z}\right) 
+ \left(1 + \frac{2}{z} \right) \ln(y_h) - \frac{1}{z} + \frac{1}{zy_h}
\Biggr]\nonumber\\ & &
~~~~~~~~~~~~~~~~~~~~\times
y_h^2~2x_m F_1(x_h,Q_h^2)\nonumber\\
& & + 2 \Biggl[\frac{1}{z} P_{\gamma e}^0(z) \left(\frac{1}{z} - y_h\right)
\ln\left(\frac{1-y_h z}{1-z} \right) + y_h \ln(y_h) +
\frac{1-y^2_h}{z} - 2 \frac{1- y_h}{z^2} \Biggr]\nonumber\\ & &
~~~~~~\times F_2(x_h,Q^2_h)\Biggr\},\\
%
%
C_{\gamma e}^{(1,1)}(y,Q^2) &=&
\frac{2\pi \alpha^2}{y_h Q_l^4} \int_{z_0^I}^1 \frac{dz}{z^2} \Biggl\{
\Biggl[\frac{1}{z} P_{e\gamma}^0(z) \ln \left(\frac{1-y_h}{1-z} \right)
+ z \ln\left(\frac{z}{y_h}\right) - 1
+ \frac{z}{y_h}\Biggr]~y^2_h~2x_m F_1(x_h,Q^2_h)\nonumber\\
& & 
~~~~~~~~~~~~~~~~
+ 2 \Biggl[\frac{1}{z^2} P_{e \gamma}^0(z) (1-y_h) \ln \left(\frac{1-y_h}
{1-z}\right) - y_h \ln\left(\frac{z}{y_h}\right) \nonumber\\
& &
~~~~~~~~~~~~~~~~
- \frac{z-y_h}{2z^2} \left(2 - y_h -3z +2y_hz\right)
\Biggr] F_2(x_h,Q^2_h) \Biggr\} \nonumber\\
& & 
+ \frac{2\pi \alpha^2}{y_h Q_l^4} \int_{z_0^F}^1 dz
\Biggl\{ \Biggl[ P_{\gamma e}^0(z) \ln\left(\frac{(1-y_h)(1-z)}
{z^2}\right) - \frac{1}{z} \ln(y_h) +2 - \frac{3}{z} + \frac{1}{y_h z}
\Biggr] \nonumber\\ & &
~~~~~~~~~~~~~~~~\times y_h^2~2 x_m F_1(x_h,Q^2_h)\nonumber\\
& & 
~~~~~~~~~~~~~~~~
+ 2 \Biggl[zP_{\gamma e}^0(z) (1-y_h) \ln\left(\frac{(1-y_h)(1-z)}
{z^2}\right) + y_h \ln(y_h) \nonumber\\ & &
~~~~~~~~~~~~~~~~-\frac{1}{2}(1-y_h)(7-y_h-14z+4y_hz+6z^2-2y_hz^2)
\Biggr]~F_2(x_h,Q^2_h) \Biggr\}. \nonumber\\
\end{eqnarray}

Finally, the term $C_{\sf v}^{(2,1)}(y,Q^2)$ is given by~:
\begin{eqnarray}
\label{eqCv}
C^{(2,1)}_{\sf v}(y,Q^2) &=&  
\int_0^1 
P_{ee,S}^{1,NS,{\rm OM}}(z) \left[\theta\left(z-z_0^I\right) J^I(z)
\widetilde{C}^{(0,0)}_I(y,Q^2) - C^{(0,0)}(y,Q^2)\right] \nonumber \\
&+ &\int_{z_0^I}^1
P_{ee,S}^{1,PS,{\rm OM}}(z)  J^I(z)
\widetilde{C}^{(0,0)}_I(y,Q^2) \nonumber\\
&+ &\int_0^1
P_{ee,T}^{1,NS,{\rm OM}}(z) \left[\theta\left(z-z_0^F\right) J^F(z)
\widetilde{C}^{(0,0)}_F(y,Q^2) - C^{(0,0)}(y,Q^2)\right] \nonumber\\
&+ &\int_{z_0^F}^1
P_{ee,T}^{1,PS,{\rm OM}}(z)  J^F(z)
\widetilde{C}^{(0,0)}_F(y,Q^2).
\end{eqnarray}
Here $P_{ee,S,T}^{1}(z)$  denote the space-- and time--like NLO QED 
splitting functions of the non--singlet $(NS)$ and pure--singlet $(PS)$ 
channels. The label OM refers to the OMS scheme. We express these
splitting functions referring to the splitting functions in the 
$\overline{\rm MS}$--scheme~\cite{NLOSP}~:
\begin{eqnarray}
\label{eqNLOSP}
P_{ee,S}^{1,NS,\overline{\rm MS}}(z)      
&=& -2P_{ee}^0(z) \ln(z) \left[\ln(1-z) + \frac{3}{4}
                       \right] - \frac{1}{2}(1+z)\ln^2(z) - \frac{1}{2}
                       (3+7z) \ln(z)  \nonumber \\
                   & & -5(1-z) -\frac{2}{3} \left\{P_{ee}^0(z) \left[\ln(z)
                       +\frac{5}{3} \right] + 2(1-z)\right\},\\
P_{ee,S}^{1,PS,\overline{\rm MS}}(z) &=&
                                -(1+z) \ln^2(z) + \left[
                       1+5z+ \frac{8}{3} z^2\right] \ln(z)
                       + \frac{2}{9} \left(\frac{1-z}{z}\right)
                       \left(10+ z +28z^2\right),
                         \\
P_{ee,T}^{1,NS,\overline{\rm MS}}(z)
&=&  2P_{ee}^0(z) \ln(z) \left[\ln(1-z) -\ln(z) + \frac{3}{4}
                       \right] + \frac{1}{2}(1+z)\ln^2(z) - \frac{1}{2}
                       (7+3z) \ln(z)  \nonumber \\
                   & & -5(1-z)-\frac{2}{3} \left\{P_{ee}^0(z) \left[\ln(z)
                       +\frac{5}{3} \right] + 2(1-z)\right\},\\
P_{ee,T}^{1,PS,\overline{\rm MS}}(z) &=&
                                  (1+z) \ln^2(z) - \left[
                       5+9z+ \frac{8}{3} z^2\right] \ln(z)
                       - \frac{4}{9} \left(\frac{1-z}{z}\right)
                       \left(5+ 23z +14z^2\right).
\end{eqnarray}
Note that the splitting functions 
$P_{ee,S,T}^{1,PS,\overline{\rm MS}}(z)$ refer to {\it one} light
fermion {\it or one} light anti--fermion, contrary to the QCD singlet
case, where always the sum of a light quark and an antiquark   is
considered. The splitting functions in the OMS are obtained by
\begin{eqnarray}
\label{SPOMS}
P_{ee,S,T}^{1,NS,{\rm OM}}(z)
&=& P_{ee,S,T}^{1,NS,\overline{\rm MS}}(z) 
+ \frac{\beta_0}{2} \Gamma_{ee}^{S,T,(1,1)}(z)~, \\
\Gamma_{ee}^{S,T,(1,1)}(z) &=& -2 \left[\frac{1+z^2}{1-z}\left(\ln(1-z)
+\frac{1}{2}\right)\right]~,
\end{eqnarray}
with $I=S, F=T$, and
$P_{ee,S,T}^{1,PS,{\rm OM}}(z) = P_{ee,S,T}^{1,PS,\overline{\rm MS}}(z)$.
\section{Conclusions}

\vspace{1mm}
\noindent
We calculated the $O(\alpha^2 L)$ leptonic QED corrections to
deep inelastic $ep$ scattering for the case of mixed variables. The
corrections are given in terms of double--differential distributions to 
be compared to the double differential Born cross section. The calculation
was performed using the renormalization--group decomposition of the
2--loop corrections to the differential cross section w.r.t. mass
factorization in the OMS scheme for the light fermion mass. Mass 
factorization introduces an artificial factorization scale $\mu^2$ on
which the physical cross section does not depend. By the elimination of 
this scale the structure of the differential cross section is 
re--organized in terms of pieces which can be calculated first 
individually and which are then composed to the desired correction term
to be obtained. The NLO correction consists out of five terms, the LO
ISR and FSR radiation correction of the non--logarithmic $O(\alpha)$ 
contribution, a respective term due to charge renormalization, two new 
terms containing $e-\gamma-e$ initial and final state transitions, 
and the ISR and FSR OMS--NLO radiation correction to the Born term.

\vspace{3mm}
\noindent
{\bf Acknowledgement.}\\ \noindent
We would like to thank W.L. van Neerven and V.
Ravindran for discussions. The work was supported in part by EU--TMR
Network HPRN-CT-2000-00149.

\end{document}